\documentclass[prb,aps,twocolumn,floatfix,amsmath,amssymb,tightenlines, showpacs, superscriptaddress]{revtex4}
\usepackage[pdftex]{graphicx}
 \usepackage{amsmath}
\usepackage{subfigure}
 \usepackage{tikz}
\usepackage[utf8x]{inputenc}
\usepackage[T1]{fontenc}
\usepackage{amssymb}
\usepackage{amsfonts}
\usepackage{bm}
 \usepackage{amsmath} 
\usepackage[breaklinks=true,colorlinks=true,linkcolor=blue,urlcolor=blue,citecolor=blue]{hyperref}
\usepackage{subfigure}
\usepackage{lipsum}
\usepackage{amsfonts} 
\usepackage{amssymb, mathrsfs}
\usepackage{braket}
\usepackage{graphicx} 
\usepackage{bbm}
\def\beq{\begin{equation}}
\def\eeq{\end{equation}}
\def\bsp{\begin{split}}
\def\esp{\end{split}}
\def\bea{\begin{eqnarray}}
\def\eea{\end{eqnarray}}
\def\ba{\begin{array}}
\def\ea{\end{array}}

\def\lb{\left(}
\def\rb{\right)}

\def\l.{\left.}
\def\r.{\right.}

\def\ra{\rangle}
\def\la{\langle}

\def\bo{\bold{k}}

\begin{document}

\title{ Quantum triangular ice in the easy-axis ferromagnetic phase}
\author{S. A. Owerre}
\email{solomon@aims.ac.za}
\affiliation{African Institute for Mathematical Sciences, 6 Melrose Road, Muizenberg, Cape Town 7945, South Africa.}
\email{sowerre@perimeterinstitute.ca}
\affiliation{Perimeter Institute for Theoretical Physics, 31 Caroline St. N., Waterloo, Ontario N2L 2Y5, Canada.}

\begin{abstract}
We use spin wave theory to investigate the ground state properties of the $Z_2$-invariant quantum XXZ model on the triangular lattice in the  ferromagnetic phase. The Hamiltonian comprises  nearest and next-nearest-neighbour Ising couplings, external magnetic fields, and a $Z_2$-invariant ferromagnetic coupling. We show that quantum fluctuations are suppressed in this system, hence linear spin wave theory gives reasonable estimates of the ground state thermodynamic properties.  Our results show that, at half-filling (zero magnetic fields), the spontaneous breaking of $Z_2$ symmetry leads to a ferromagnetic phase whose energy spectrum is gapped at all excitations with a {\it maxon} dispersion at $\bold{k}=0$. This is in sharp contrast to  rotational invariant systems with a vanishing {\it phonon} dispersion.  We show that the $\bo=0$ mode enhances the estimated values of the thermodynamic quantities.  We obtain the trends of the particle density and  the condensate fraction. The density of states and the dynamical structure factors exhibit fascinating peaks at unusual wave vectors, which should be of interest.
\end{abstract}

\pacs{75.10.Jm, 75.30.Ds, 05.30.Jp, 75.40.Gb}

\maketitle

\textbf {Introduction}--. Quantum spin ice (QSI) on three-dimensional pyrochlore lattice has become the subject of intense research in recent years. \cite{gin, zhi, sun4a} In this system, the spin ordering at the vertices of the pyrochlore lattice, that is two-in, two-out, is reminiscent of hydrogen atoms in  water ice.  Competing interactions between frustrated spins give rise to fascinating physics with rich quantum phases.  Huang, Chan and Hermele\cite{Huang} recently proposed an alternative simplified Hamiltonian that captures the physics of QSI on three-dimensional pyrochlore lattice. In the presence of a [111] crystallographic field, the spin configurations exhibit an ordering which can be mapped onto a two-dimensional kagome lattice.\cite{juan} The Hamiltonian, however, retains the same form as the Huang, Chan and Hermele model.   Recently, Carrasquilla {\it et~al.,}~\cite{juan} have uncovered the phase diagram of this model on the kagome lattice by Quantum Monte Carlo (QMC) simulations.  The Hamiltonian studied by Carrasquilla {\it et~al.,}~ \cite{juan} is termed {\it Quantum Kagome Ice} and it has the form
\begin{align}
&H= -\frac{J_{\pm\pm}}{2}\sum_{\la lm\ra}\lb  S_{l}^+S_{m}^+ +S_{l}^-S_{m}^-\rb+ J_z\sum_{\la lm\ra} S_{l}^zS_{m}^z\label{k1}\nonumber\\&-h_z\sum_l S_l^z, 
\end{align}
where $S_{l}^{\pm}= S_{l}^{x} \pm  i S_{l}^{y}$ are the raising and the lowering spin operators respectively. In accordance with previous terminology, the term {\it Quantum Triangular Ice} refers to the study of Eq.~\eqref{k1} on the triangular lattice.   Here, $J_z>0$ is the frustrated nearest-neighbour (nn) Ising coupling, and $J_{\pm\pm}>0$ is the unfrustrated $Z_2$-invariant easy-axis ferromagnetic coupling. The sign of $J_{\pm\pm}$ is immaterial by virtue of the unitary transformation $S_{lm}^\pm\to \pm i S_{lm}^\pm$.  

The crucial difference between the  U(1)-invariant XXZ model \cite{AW, AW1, kle1,isaa, isaa1, isaa2,  isa} and  Eq.~\eqref{k1} is that the ferromagnetic interaction in the former is easy-plane  and has a unique sign for non-bipartite lattices, whereas in the latter, the ferromagnetic interaction is easy-axis and the sign is immaterial, which results in a $Z_2$-invariant Hamiltonian in the $x$-$y$ plane. Thus,  Eq.~\eqref{k1}  cannot be mapped onto a  U(1)-invariant XXZ model (hard-core boson) on non-bipartite lattices, whereas for bipartite lattices, Eq.~\eqref{k1} is related to the U(1)-invariant counterpart by a simple spin flip.   QMC simulations of  Carrasquilla {\it et.~al}~\cite{juan} on the kagome lattice uncovered  three distinct phases, which include the unconventional disordered magnetized lobes for small $\pm h_z$ and $J_{\pm\pm}/J_z<0.5$, and it is believed to be a candidate for Quantum Spin Liquid (QSL) state, enabling the  search for two-dimensional QSL within a class of pyrochlore QSI materials.  The remaining two phases are the $S_x$ and the $S_z$ ferromagnetic ordered phases for small and large $h_z$ respectively. 

We have provided an analytical explanation of the QMC results using spin wave theory on the kagome lattice.\cite{sow1} Remarkably, we observed  that the average spin-deviation operator  $\la n_l\ra=\la b_l^\dagger b_l\ra$ is very small in {\it Quantum Kagome Ice}, {\it i.e.,} quantum fluctuations are suppressed.  Hence, we were able to capture a broad trend of the quantum phases  uncovered by QMC simulations.\cite{juan}  We have also obtained the complete phase diagram of Eq.~\eqref{k1}  on the triangular lattice ({\it Quantum Triangular  Ice}), using the semiclassical large-$S$ expansion. \cite{sow2}   In the case of triangular lattice, there is an additional phase at  $h_z=0$ and $J_z\to \infty$. This phase is selected by quantum fluctuations via order-by-disorder mechanism and it is called a {\it ferrosolid} state --- a state similar to a  {\it supersolid} state in the  U(1)-invariant XXZ model,\cite{ kle1,isaa, isaa1, isaa2,  isa} but exhibits broken translational and $Z_2$ symmetries. It also appears adjacent to the  unconventional disordered magnetized lobes for small $\pm h_z$. Again, spin wave theory provides an accurate picture of the quantum phases. The  XY limit of this model on both lattices has also  been analyzed using spin wave theory and QMC.\cite{sow3}

In this paper, we explore the ground state properties of the easy-axis ferromagnetic phase of this model on the triangular lattice. A related U(1)-invariant XXZ model in the superfluid phase has been studied previously on the triangular lattice, using series expansion methods\cite{AW} and spin wave theory. \cite{AW1} The present model has not been studied in the ferromagnetic phase. The study of this model has some physical relevance  since many magnetic materials are ferromagnets. This includes the unconventional CeRh$_3$B$_2$, which has been studied by many authors.\cite{sr} The ferromagnetic XY coupling also plays a prominent role in the experimental realization  of hard-core bosons in ultracold atoms on quantum optical lattices.\cite{bec,mar} Hence, it is expedient to understand the ferromagnetic nature of the $Z_2$-invariant model, as it  might be applicable to these systems. We will consider a more general Hamiltonian given by
\begin{align}
&H=-\frac{J_{\pm\pm}}{2}\sum_{\la lm\ra} \lb S_{l}^+S_{m}^+ +S_{l}^-S_{m}^-\rb + \sum_{lm}J_{l,m}  S_{l}^zS_{m}^z \nonumber\\&-h_z\sum_l S_l^z-h_x\sum_l S_l^x,
\label{kk}
\end{align}
where $J_{l,m}=J_z$ on the nn sites and $J_{l,m}=J^{\prime}_z$ on the next-nearest-neighbour (nnn) sites.  The external magnetic fields are introduced to enable the calculation of parallel and longitudinal magnetizations. The $Z_2$ symmetry of Eq.~\eqref{kk} at $h_x=0$ is synonymous with the fact that  the total $S_z$ is not conserved.  
 
  In this model, spin wave theory is exact at the Heisenberg point $J_z^\prime=h_{x,z}=0$ and $J_z=J_{\pm\pm}=J$. The exact ground state in this limit is a ferromagnet along the $x$-direction.  The excitation spectrum, however,  exhibits no soft (Goldstone) modes at $\bo=0$ and the average spin-deviation is not divergent at finite temperature. Hence, the discrete $Z_2$ symmetry is spontaneously broken even at finite temperature, thus  Mermin-Wagner theorem\cite{mer} is not applicable. This unusual feature is obviously  absent in rotational invariant systems.  Away from the Heisenberg point, the average spin-deviation $\la n_l\ra$ does not exceed $0.025$ for all parameter regimes considered at half-filling $h_{x,z}=0$. Thus, spin wave theory gives a very good description of the system. We calculate the excitation spectrum $\epsilon(\bo)$, the density of states, and the static structure factors $\mathcal{S}^{zz}(\bo)$ and $\mathcal{S}^{\pm}(\bo)$.  We also calculate the particle density and the non-divergent condensate fraction at $\bo=0$. In contrast to the U(1)-invariant model, with a phonon dispersion near $\bo=0$, \cite{AW,AW1}   the spectrum of the $Z_2$-invariant model exhibits a {\it maxon}  dispersion near $\bo=0$ with a gap  of $\Delta\propto \sqrt{J_{\pm\pm}(J_z+J_z^\prime +J_{\pm\pm})}$ at half-filling ($h_{x,z}=0$). We see that the gap does not vanish for  $J_{\pm\pm}\neq 0$.   For fixed $J_z^\prime, J_{\pm\pm}$ and varying $J_z$, we observe well-defined roton minima at some points inside the Brillouin zone. The spectrum also exhibits some profound flat mode at some paths of the Brillouin zone and the gap at the Brillouin zone corners $\bold{K}=(\pm 4\pi/3, 0)$ vanishes at  $J_z^\prime=J_{\pm\pm}-J_z/2$. Thus, for $J_z^\prime=0$ the transition between the  easy-axis ferromagnet and the {\it ferrosolid} occurs at $J_z=2J_{\pm\pm}$, which recovers the result obtained in the U(1)-invariant XXZ model.\cite{kle1} However, spin wave theory gives a better description of the present model.  The static structure factors exhibit some distinctive features with sharp peaks at the minima of the energy spectra and no discontinuity in the entire Brillouin  zone. Likewise, the density of states exhibit interesting features with peaks at various energies.

\textbf{Linear spin wave theory}--.
\label{class}
We  now consider the anisotropic Hamiltonian in Eq.~\eqref{kk} in the easy-axis ferromagnetic phase.   At the Heisenberg point $J_z^\prime=h_{x,z}=0$ and $J_z=J_{\pm\pm}=J$,  the resulting Hamiltonian  can be written as
\bea
H=J\sum_{\la lm \ra} -S_{l}^xS_{m}^x +\frac{1}{2}(S_{l}^+S_{m}^- + S_{l}^-S_{m}^+).
\label{spl}
\eea
We have chosen $x$-quantization axis, hence $S_{l}^\pm = S_l^z\pm iS_l^y$. For bipartite lattices, Eq.~\eqref{spl} can be transformed to  SU(2) Heisenberg ferromagnet by a $\pi$-rotation about the $x$-axis on one sublattice, {\it i.e.},  $S_m^x\to S_m^x$,  $S_m^\pm\to -S_m^\pm$. For non-bipartite lattices, such rotation cannot be performed, thus Eq.~\eqref{spl} differs from the SU(2)-invariant Heisenberg ferromagnet, and exhibits   $Z_2\times$U(1) symmetry with $U(1)$ symmetry  in the $z$-$y$ plane and $Z_2$ symmetry in the $x$-axis. The  easy-axis ferromagnetic ordered state results from the spontaneous  breaking of $Z_2$ symmetry along the $x$-axis, $\la S_x\ra\neq 0$ and $\la S_{zy}\ra=0$. Thus, for spin-$1/2$, the state $\ket{\psi_{FM}}=\prod_l\ket{S_{l}^x=\uparrow}$  is an exact eigenstate of Eq.~\eqref{spl} with $\mathcal{E}_{MF}=-3JNS^2$,  and $\la S_x\ra=S$. As we will show, the resulting excitation above this ground state has  a gapped quadratic dispersing mode near $\bo=0$, signifying  no Goldstone mode due to the absence of a complete continuous symmetry. This is in sharp contrast to the superfluid phase with  broken $U(1)$ symmetry in the $x$-$y$ plane $\la S_{xy}\ra\neq 0$ and a gapless linear excitation at $\bo=0$. 

For the anisotropic model, the mean field energy in the ferromagnetic phase is given by
\begin{align}
&\mathcal{E}_{MF}(\theta)= 3{N}S^2( \lambda_{1z}+ \lambda_{2z}-{h}/{3}),
\label{emf}
\end{align}
where $N$ is the total number of sites and
\begin{align}
&\lambda_{1z}= J_z\cos^2\theta-J_{\pm\pm}\sin^2\theta;\quad\lambda_{2z}= J_z^\prime\cos^2\theta;
\\&h=h_z\cos\theta +h_x\sin\theta. 
\end{align}
The minimization of the mean field energy with respect to $\theta$ yields
\begin{align}
h_z-6\mathcal J \cos\vartheta=h_x\cot\vartheta,
\label{hh1}
\end{align}
where $\vartheta$ is the angle that minimizes Eq.~\eqref{emf} and $\mathcal J=\lb J_z+J_z^\prime+J_{\pm\pm} \rb$. We can eliminate $h_x$ from Eq.~\eqref{emf} using Eq.~\eqref{hh1},  the classical energy becomes
\begin{align}
\mathcal{E}_{MF}(\vartheta)= 3NS^2[\mathcal J \sin^2\vartheta + J_z+J_z^\prime-h_z\sec\vartheta/3].
\end{align}
It should be noted that $\vartheta$ is a function of $h_x$. 
At $h_x=0$, the solution of Eq.~\eqref{hh1} gives $\cos\vartheta_0=h_z/h_c$, where $h_c=6\mathcal J$ is the critical field above which the spins  are fully polarized along the $z$-direction. The corresponding classical energy is given by
\begin{align}
\mathcal{E}_{MF}(\vartheta_0)= -3{N}S^2[\mathcal J \cos^2\vartheta_0+J_{\pm\pm}].\label{mf2} 
\end{align}
For $h_x\neq 0$, the mean field energy is obtained perturbatively for small $h_x$,
\begin{align}
\mathcal{E}_{MF}(\vartheta)=\mathcal{E}_{MF}(h_x=0)+ h_x\left.\frac{\partial \mathcal{E}_{MF}(\vartheta)}{\partial h_x}\right\vert_{h_x=0} +\cdots,
\end{align}
where $\mathcal{E}_{MF}(h_x=0)=\mathcal{E}_{MF}(\vartheta=\vartheta_0)$.
We perform  spin wave theory in the usual way, by rotating the coordinate about the $y$-axis in order to align the spins along the selected direction of the magnetization. 
\begin{align}
&S_l^x=S_l^{\prime x}\cos\theta  +  S_l^{\prime z}\sin\theta,\label{trans}\nonumber\\&
S_l^y=S_l^{\prime y},\\&\nonumber
S_l^z=- S_l^{\prime x}\sin\theta + S_l^{\prime z}\cos\theta.
\end{align}
Next, we express the rotated coordinates in terms of the linearized Holstein Primakoff (HP) transform.

 \begin{align}
 &S_{l}^{\prime z}= S-b_{l}^\dagger b_{ l}, \label{HP}\nonumber\\&
 S_{l}^{\prime y}= i\sqrt{\frac{S}{2}}\lb b_{l}^\dagger -b_{l}\rb,
 \\&\nonumber
 S_{l}^{\prime x}= \sqrt{\frac{S}{2}}\lb b_{l}^\dagger +b_{l}\rb.
 \end{align}
 The truncation of the HP transformation at linear order is guaranteed provided the average spin-deviation operator $\la n_l\ra=\la b_l^\dagger b_l\ra$ is small. Indeed,  $\la n_l\ra$ is small in the present model, hence linear spin wave theory is suitable  for the description of this system.
Taking the magnetic fields, $h_{x,z}$, to be of order $S$ and keeping only the quadratic terms,
 the resulting bosonic Hamiltonian can be diagonalized by the Bogoliubov transformation,
\begin{align}
b_{\bo}=u_{\bo}\alpha_{\bo}-v_{\bo}\alpha_{-\bo}^\dagger,
\end{align}
where $u_{\bo}^2-v_{\bo}^2=1$, one finds that the resulting Hamiltonian is diagonalized by
\begin{align}
&u_{\bo}^2=\frac{1}{2}\lb \frac{A_{\bo}(\vartheta)}{\omega_{\bo}(\vartheta)}+1\rb; \quad v_{\bo}^2=\frac{1}{2}\lb \frac{A_{\bo}(\vartheta)}{\omega_{\bo}(\vartheta)}-1\rb,\end{align}
with $\omega_\bo(\vartheta)=\sqrt{A_\bo^2(\vartheta)-B_\bo^2(\vartheta)}$.  The diagonal Hamiltonian yields
\bea
H=S\sum_{\bo}\omega_{\bo} (\vartheta)\lb \alpha_{\bo}^\dagger \alpha_{\bo}+\alpha_{-\bo}^\dagger \alpha_{-\bo}\rb.
\eea

The excitation of the quasiparticles is given by

\bea
\epsilon_\bo(\vartheta)=2S\omega_\bo(\vartheta)=2S\sqrt{A_\bo^2(\vartheta)-B_\bo^2(\vartheta)},
\eea
while the spin wave ground state energy is given by \bea \mathcal{E}_{SW}(\vartheta)=\mathcal{E}_{MF}(\vartheta) +S\sum_{\bo}[\omega_{\bo}(\vartheta)-\mathcal{E}_{lo}(\vartheta)] \label{Esw},\eea
where
\begin{align}
&A_\bo(\vartheta)=\mathcal{E}_{lo}(\vartheta) +3J_{\pm\pm}\gamma_{\bo}+ B_\bo(\vartheta);\\& B_\bo(\vartheta)=\frac{3}{2}\lb \bar{g}_\bo(\vartheta)+g_\bo(\vartheta)\rb;
\end{align}

\begin{align}
&\mathcal{E}_{lo}(\vartheta)=-3(J_{z}+J_{z}^\prime)+h_z\sec\vartheta/2,\\&
\bar{g}_\bo(\vartheta)=(J_z+J_{\pm\pm})\sin^2\vartheta\gamma_{\bo}-2J_{\pm\pm}\gamma_{\bo},\\&g_\bo(\vartheta)=J_z^\prime\sin^2\vartheta\bar{\gamma}_\bo.
\end{align}
We have eliminated $h_x$ using Eq.~\eqref{hh1}.
The structure factors are given by
\begin{align}
&\gamma_{\bo}=\frac{1}{3}\lb \cos k_x+2\cos \frac{k_x}{2}\cos \frac{\sqrt{3}k_y}{2}\rb,\\&
\bar{\gamma}_{\bo}=\frac{1}{3}\lb \cos \sqrt{3}k_y+2\cos \frac{3k_x}{2}\cos \frac{\sqrt{3}k_y}{2}\rb.
\end{align}

\begin{figure}[ht]
\centering
\includegraphics[width=3in]{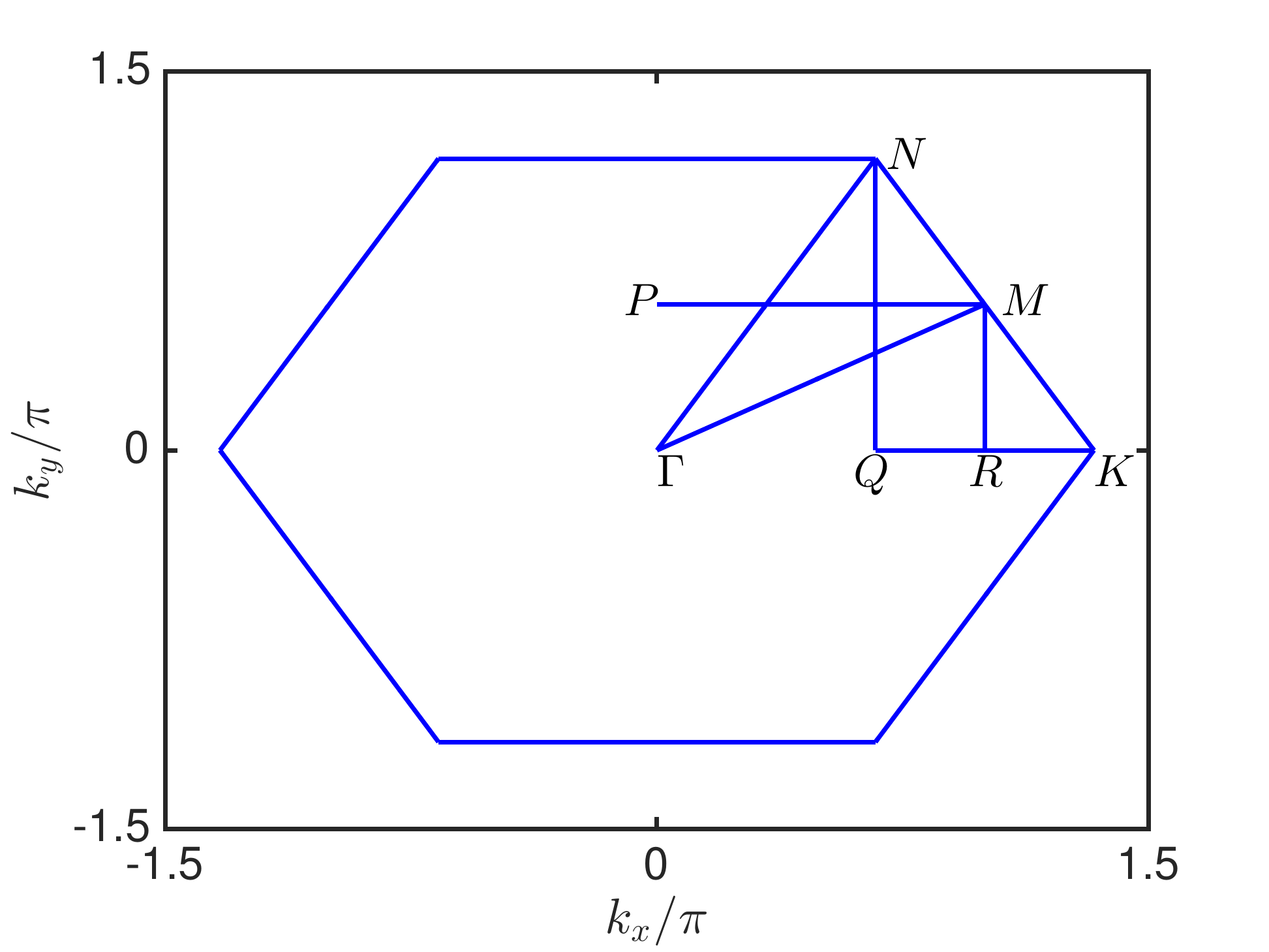}
\caption{Color online.  The Brillouin zone of the triangular lattice and the corresponding paths that will be adopted in this paper.}
\label{K_spec1}
\end{figure}

\begin{figure}[ht]
\centering
\includegraphics[width=3in]{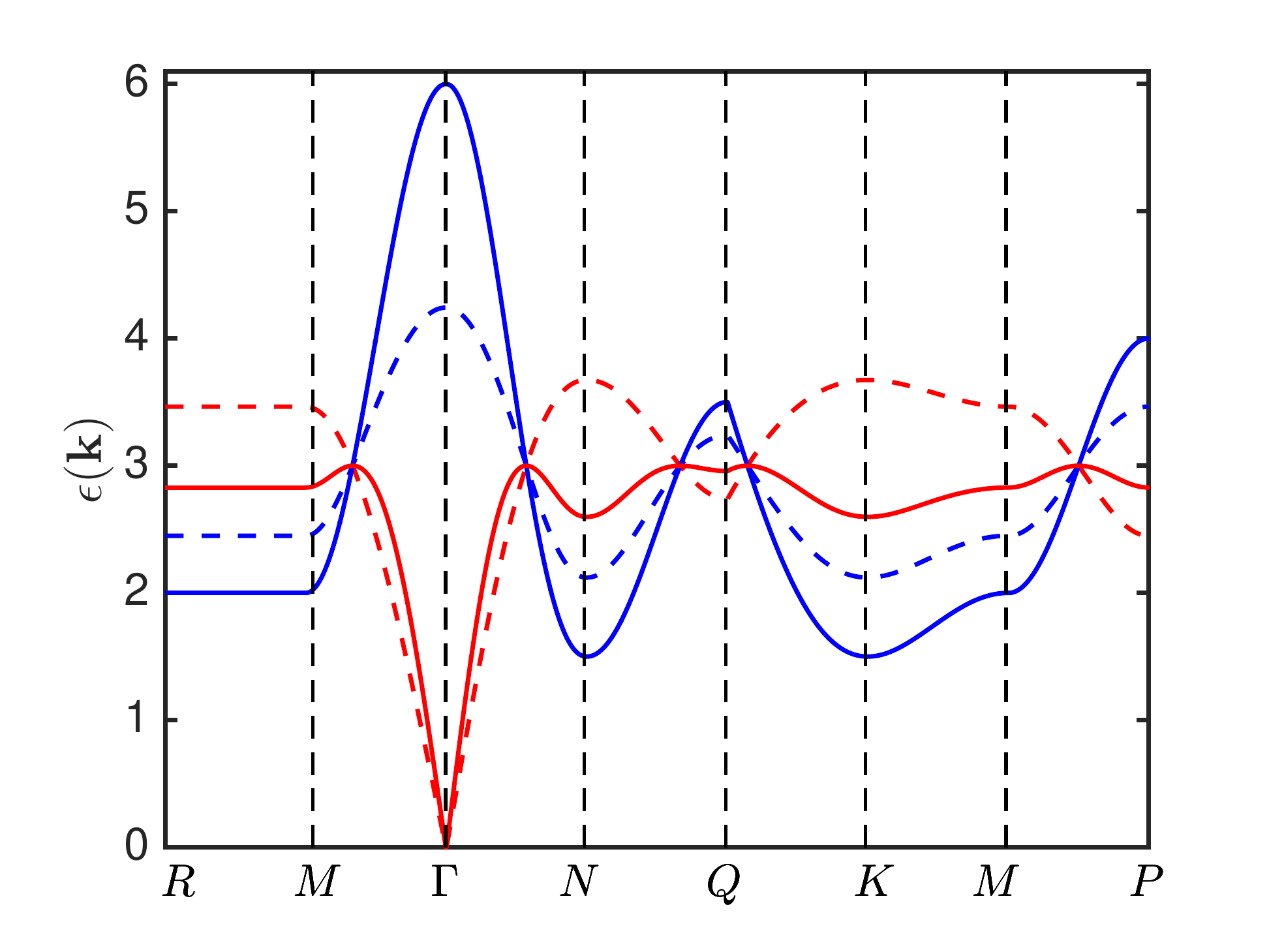}
\caption{Color online. The plots of the energy dispersion  at $h_{x,z}=0~(\rho=0.5)$ and $J_{\pm\pm}=1;~ J_z^\prime=0$. XY model: $J_z=0$  (dashed). Heisenberg model:  $J_z =1$ (solid). The blue curves denote the present model and the red curves denote the U(1) model.}
\label{ee_nn0}
\end{figure}

\textbf{Excitation spectra}--. We now investigate the nature of the energy spectra of Eq.~\eqref{kk}. We will be interested in half-filling or zero magnetic fields  and spin-$1/2$. We adopt the Brillouin zone paths of Refs.~[\onlinecite{AW, AW1}] as shown in Fig.~\eqref{K_spec1}. Our main focus is the appearance of a minimum inside the Brillouin zone (roton minimum) and the possibility of any soft modes. The vanishing of the spectrum at the corners of the Brillouin zone represents a phase transition to a new spin configuration.  The simplest case is the Heisenberg point  $h_{x,z}=J_z^\prime=0;~ J_z=J_{\pm\pm}=J$. In this limit spin wave theory is exact  and the excitation spectrum of Eq.~\eqref{spl} is $\omega_\bo=A_\bo= 3J(1+\gamma_\bo)$, where  $B_\bo=0$. We see that the exact ground state is the fully polarized easy-axis ferromagnet along the $x$-axis, and the corresponding excitation exhibits no zero modes since $-1/2\leq \gamma_\bo\leq 1$; see Fig.~\eqref{ee_nn0}. Near $\bo=0$, the energy behaves as $\omega_\bo\approx a-b\bo^2$, with $a=6J$ and $b=3J/4$. Hence,  the average spin deviation operator near this mode at low-temperature is given by
\begin{align}
\Delta S_x(T)=\int \frac{ d\bo}{e^{\omega_\bo/k_BT}-1}\sim~ T\frac{\ln(1-e^{-a})}{b}.
\label{mw}
\end{align}
This unusual feature means that the discrete $Z_2$  symmetry is spontaneously broken even at finite temperature, thus   Mermin-Wagner theorem\cite{mer} does not apply.
 Another important feature of Eq.~\eqref{mw} is that we can suppress quantum fluctuations by making the gap ``$a$'' as large as possible. 

\begin{figure}[ht]
\centering
\includegraphics[width=3in]{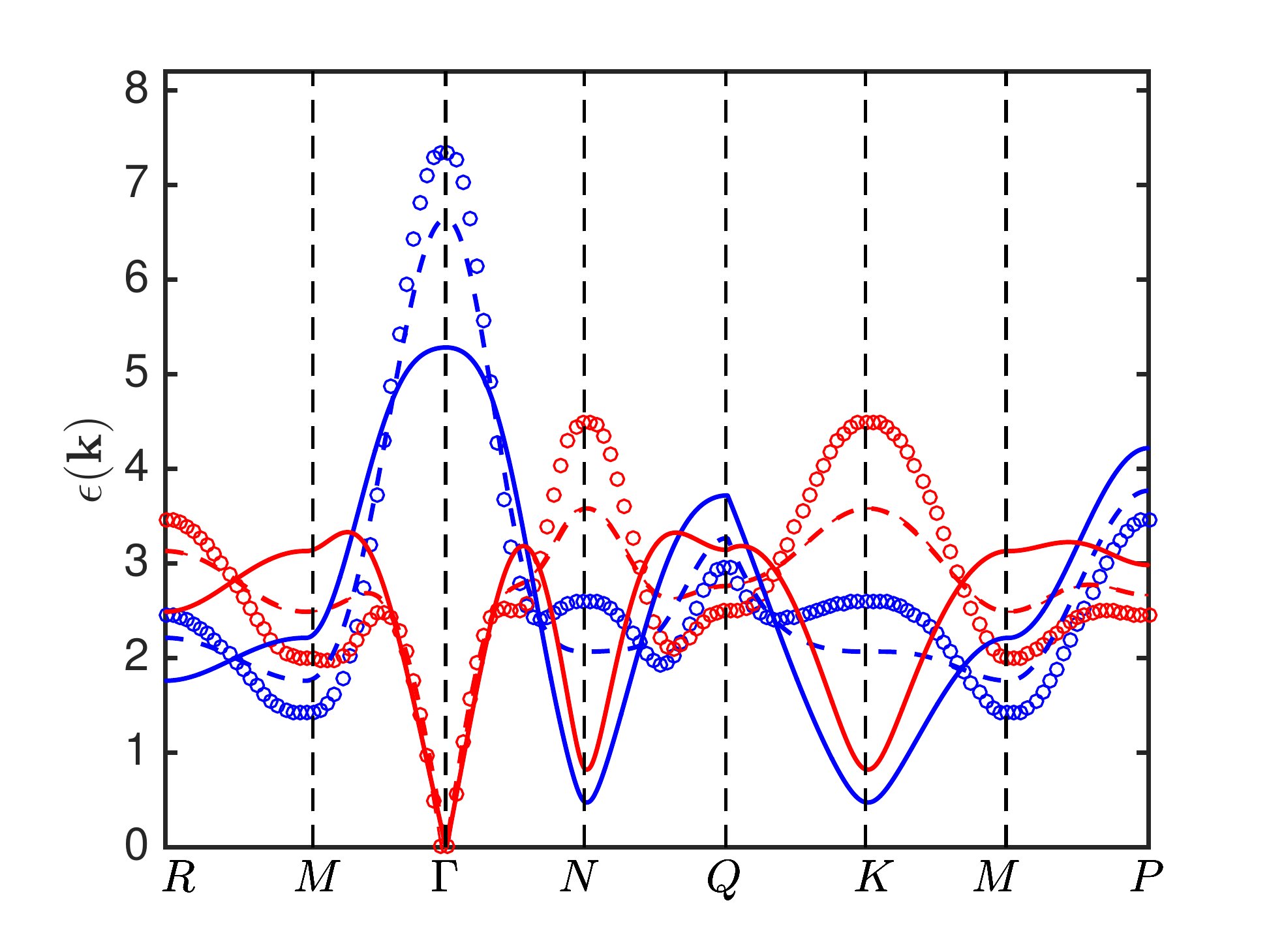}
\caption{Color online. The plots of the energy dispersion  at $h_{x,z}=0~(\rho=0.5)$, $J_{\pm\pm}=J_z=1$ and several values of $J_z^\prime=0.45$ (dashed),  $J_z^\prime =-0.45$ (solid), and  $J_z ^\prime =1$ (symbol). The colors have the same meaning as in Fig.~\eqref{ee_nn0}.}
\label{ee1}
\end{figure}
 As depicted in Figs.~\eqref{ee_nn0} and \eqref{ee1}, the energy spectra of the $Z_2$-invariant XXZ model have a similar behaviour to the U(1)-invariant XXZ model\cite{AW, AW1} at the corners of the Brillouin zone and along $RM$ except for the case of XY model. At the corners of the Brillouin zone,  the spectrum vanishes when $J_z^\prime=J_{\pm\pm}-J_z/2$. Hence at $J_z^\prime=0$, the transition from the easy-axis ferromagnet to a {\it ferrosolid} occurs at $J_{\pm\pm}-J_z/2=0$, which happens to be the same as the U(1)-invariant model.\cite{kle1}  As we will show in the subsequent  section, spin wave theory gives a better description of the present model.  In contrast to the U(1)-invariant model, the spectra of the $Z_2$-invariant model is  invariably gapped in the entire Brillouin zone and the spectra display a maxon dispersion at the $\Gamma$ point ($\bo=0$) as opposed to the usual phonon dispersion in rotational invariant systems.  The gapped nature of the $Z_2$-invariant model is as a consequence of the $Z_2$-symmetry of the Hamiltonian and it plays a very crucial role in the quantum phase diagram of the {\it Quantum Kagome Ice} uncovered by QMC.\cite{juan,sow1} The contribution from the $\bo=0$ mode is the most important feature of the $Z_2$-invariant model.   The gap at the $\Gamma$ point behaves as $\Delta\propto \sqrt{J_{\pm\pm}(J_z+J_z^\prime +J_{\pm\pm})}$, which vanishes only for $J_{\pm\pm}=0$.   In the subsequent  section, we will show that the $\bo=0$ mode enhances the estimated values of the thermodynamic quantities.  These trends are slightly  modified away from half-filling.

\textbf{Thermodynamic quantities}--. The effects of the gapped excitations in the preceding section are manifested explicitly in the  ground state thermodynamic quantities. We compute  the magnetizations per site given by\cite{ ber, tom}
\begin{align}
&\la S_z\ra= -\frac{1}{S N}\frac{\partial \mathcal{E}_{SW}(\vartheta_0)}{\partial  h_z},\label{sz}\\&
\la S_x\ra= -\frac{1}{S N}\left.\frac{\partial \mathcal{E}_{SW}(\vartheta)}{\partial  h_x}\right\vert_{\vartheta=\vartheta_0}\label{sx}.
\end{align}
Using Eq.~\eqref{Esw} we obtain
 \begin{align}
&\la S_z\ra= S\cos\vartheta_0 + \frac{\cos\vartheta_0}{2\Theta_{0}}\frac{1}{N}\sum_{\bo}\Theta_{\bo}\sqrt{\frac{{A}_ {\bo}(\vartheta_0)-{B}_ {\bo}(\vartheta_0)}{A_ {\bo}(\vartheta_0)+{B}_ {\bo}(\vartheta_0)}}.
\label{sz1}
\end{align}
with $\Theta_\bo = (J_z+J_{\pm\pm})\gamma_\bo + J_z^\prime \bar{\gamma}_\bo$. 
The total density of particles is given by $\rho= S+\la S_z\ra$.
To linear order in $h_x$ we find
\begin{align}
&\la S_x\ra= S\sin\vartheta_0 - \frac{\cos^2\vartheta_0}{2\Theta_0\sin\vartheta_0}\frac{1}{{N}}\sum_{\bo}\Theta_{\bo}\sqrt{\frac{{A}_ {\bo}(\vartheta_0)-{B}_ {\bo}(\vartheta_0)}{{A}_ {\bo}(\vartheta_0)+{B}_ {\bo}(\vartheta_0)}}\label{sx1}\\&\nonumber
-\frac{1}{2\sin\vartheta_0}\frac{1}{{N}}\sum_{\bo}\bigg[\frac{{A}_ {\bo}(\vartheta_0)}{\omega_ {\bo}(\vartheta_0)}-1\bigg].
\end{align}
\begin{figure}[ht]
\centering
\includegraphics[width=3in]{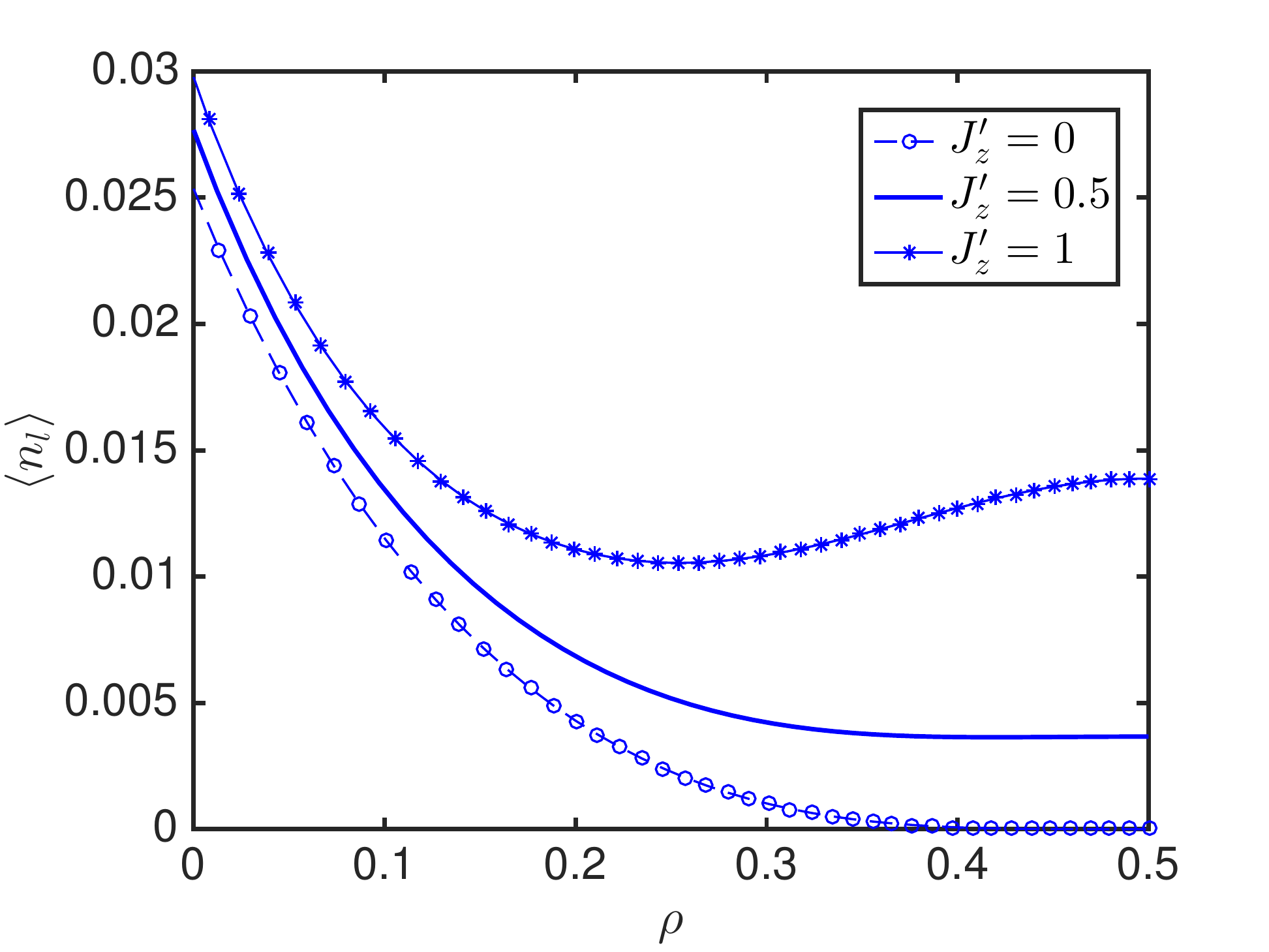}
\caption{Color online. The spin-deviation operator against the particle density at $h_{x}=0$, $J_{\pm\pm}=J_z=1$. }
\label{ave}
\end{figure}
\begin{figure}[ht]
\centering
\includegraphics[width=3in]{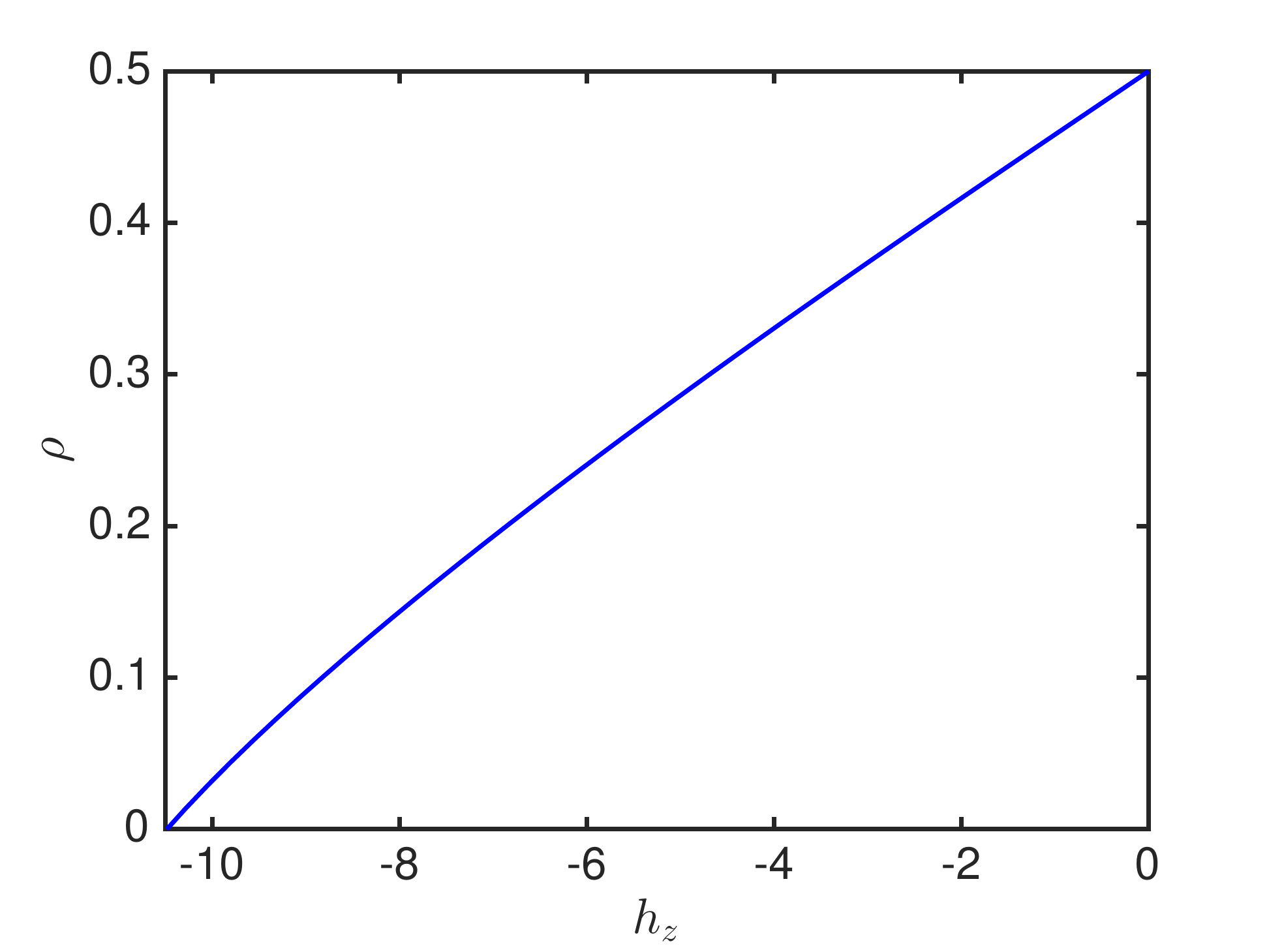}
\caption{Color online. The particle density $\rho$ vs. $h_z$ at $J_z=J_{\pm\pm}=1$, $J_z^\prime=0$, and $S=1/2$.}
\label{rho_hz}
\end{figure} 
Similar to the U(1)-invariant model, the condensate fraction at $\bo=0$ is related to the $S_x$-order parameter by $\rho_0=\lim_{S\to1/2}\la S_x\ra^2$. To linear order in spin wave theory $\rho_0$ is given by
\begin{align}
&\rho_0= \rho_0^c- \frac{\cos^2\vartheta_0}{2\Theta_0}\frac{1}{{N}}\sum_{\bo}\Theta_{\bo}\sqrt{\frac{{A}_ {\bo}(\vartheta_0)-{B}_ {\bo}(\vartheta_0)}{{A}_ {\bo}(\vartheta_0)+{B}_ {\bo}(\vartheta_0)}}\label{sx1}\\&\nonumber
-\frac{1}{2}\frac{1}{{N}}\sum_{\bo}\bigg[\frac{{A}_ {\bo}(\vartheta_0)}{\omega_ {\bo}(\vartheta_0)}-1\bigg].
\end{align}
where $\rho_0^c =\frac{1}{4}\sin^2\vartheta_0$ is the classical condensate fraction.
\begin{figure}[ht]
\centering
\includegraphics[width=3in]{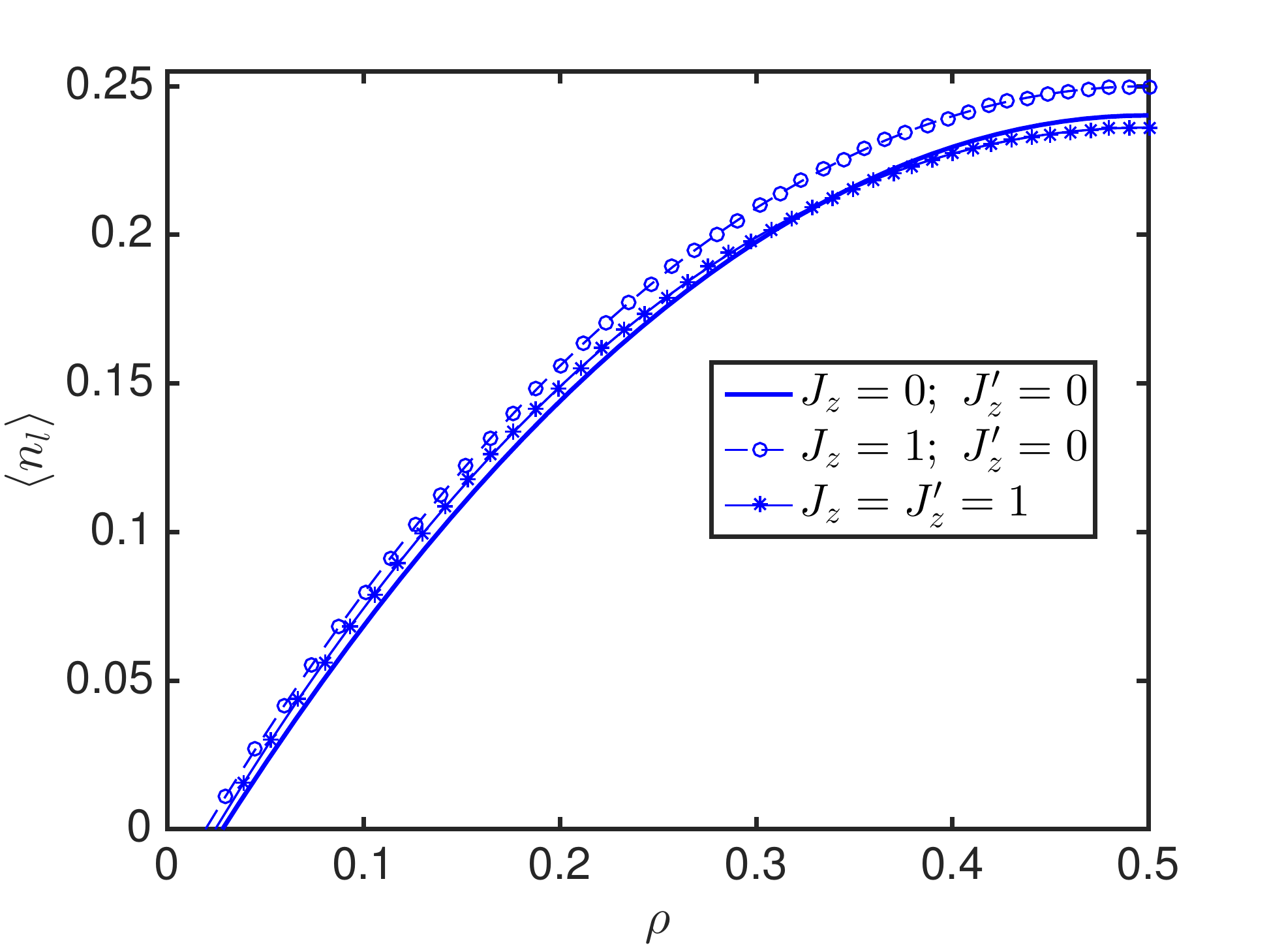}
\caption{Color online. The condensate fraction, $\rho_0$, against the particle density $\rho$ at  $J_{\pm\pm}=1$ and $S=1/2$.}
\label{rho0}
\end{figure}
 An important feature of the $Z_2$-invariant model is that all the thermodynamic quantities are finite at all points in the Brillouin zone. This enhances the estimated values of the thermodynamic quantities.  At the XY point,  $J_z=J_z^\prime=h_{x,z}=0$, the estimated value of the order parameter is $\la S_x\ra= S-0.00981$,\cite{sow3} which should be compared to the O(2)-invariant model $\la  S_x\ra= S-0.05146$.\cite{zhe}  We see that quantum fluctuations are suppressed in the $Z_2$-invariant model. A detail analysis of the XY model has been given elsewhere for the triangular and the kagome lattices.\cite{sow3} As we can see from Fig.~\eqref{ave},  the spin-deviation operator $\la n_l\ra$  is extremely small close to half-filling and also very small away from half-filling. In other words, linear spin theory is very suitable for describing the ground state properties of the $Z_2$-invariant XXZ model on non-bipartite lattices.\cite{sow1, sow2, sow3} We have shown the trend of the particle density in Fig.~\eqref{rho_hz} as a function of $h_z$.    In Fig.~\eqref{rho0}, we plot the condensate fraction at $\bo=0$ as a function of $\rho$.  At $\rho=0.5$, the estimated values of $\rho_0$ at the XY point are $\rho_0=0.2402$ linear spin wave theory of the $Z_2$-invariant model,\cite{sow3} $\rho_0=0.2011$ series expansion of the O(2)-invariant model,\cite{zhe} and $\rho_0=0.19127$ second-order spin wave theory on the square lattice. \cite{tom} At the  Heisenberg point,  $\rho_0=\rho_0^c=0.25$ as expected.  
%

\textbf{Structure factors}--. 
Due to the gapped nature of the $Z_2$-invariant model, all the thermodynamic quantities behave nicely without any divergent contributions. The density of states for this model is depicted in Fig.~\eqref{DOS} for several values of the anisotropies. There are some striking  features in the density of states of this model. We observe several spikes (van Hove singularities) depending on the anisotropies, which stem from the $\bo=0$ mode and the flat mode along $RM$. In the case of XY model, the major contribution to the  spike comes from the  $\bo=0$ mode and the discontinuity is a result of the lowest energy states at the roton minima  at the corners of the Brillouin zone.

Let us turn to the calculation of the dynamical structure factors, which is given by 
\begin{align}
\mathcal{S}^{\beta\gamma}(\bo,\omega)= \frac{1}{2\pi}\int_{-\infty}^{\infty}dt ~e^{i\omega t}\la S^\beta _{\bo}(t)S^\gamma_{-\bo}(0)\ra,
\end{align}
where $
S^\beta_{\bo}= \frac{1}{\sqrt N}\sum_l e^{-i\bo l}S_{l}^\beta
$ is the Fourier transform of the operators, and $\beta,\gamma=(x,y,z)$ label the components of the spins.  At zero temperature, the structure factors are obtained from the Green's function:
\begin{figure}[ht]
\centering
\includegraphics[width=3in]{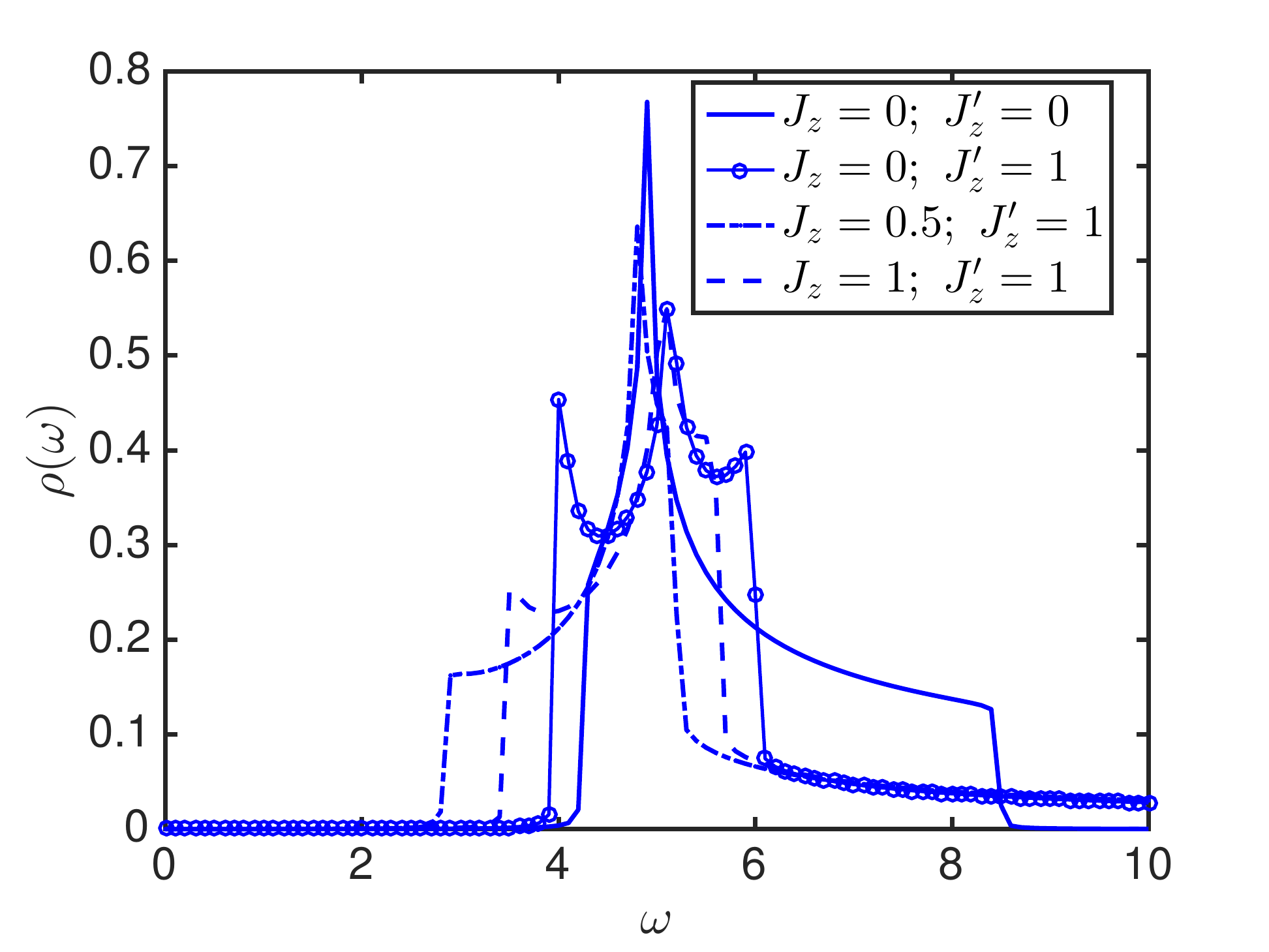}
\caption{Color online. The density of states $D(\omega)$ vs. $\omega$ at half-filling, $h_{x,z}=0~(\rho=0.5)$; $J_{\pm\pm}=J_z^\prime=1$.}
\label{DOS}
\end{figure}
\begin{figure}[ht]
\centering
\includegraphics[width=3in]{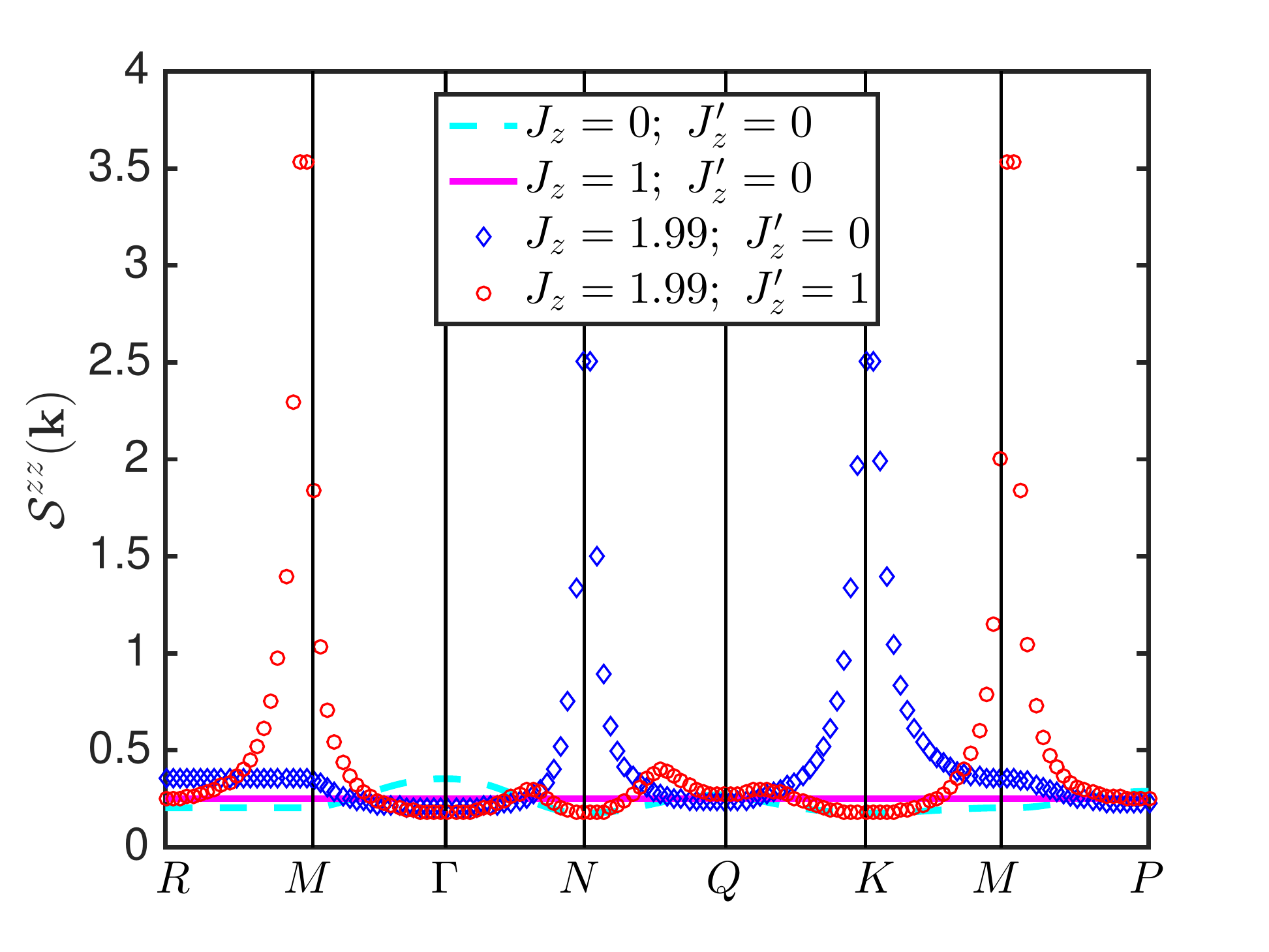}
\caption{Color online. The static dynamical structure factor, $\mathcal{S}^{zz}(\bo)$, along the Brillouin zone paths in Fig.~\eqref{K_spec1} at $h_{x,z}=0~(\rho=0.5)$;  $J_{\pm\pm}=1$ and $S=1/2$.}
\label{Tsf}
\end{figure} 
\begin{figure}[ht]
\centering
\includegraphics[width=3in]{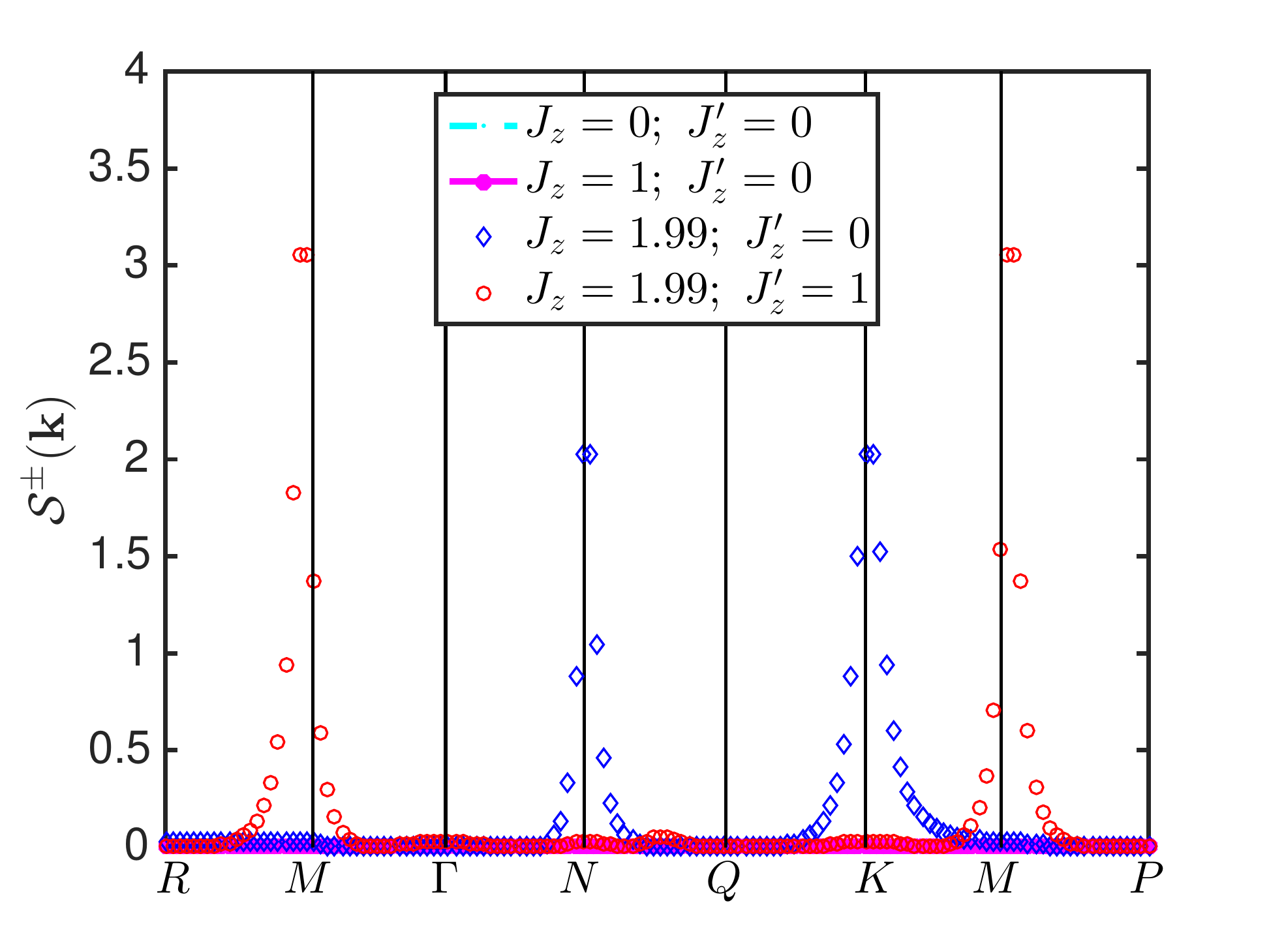}
\caption{Color online. The static dynamical structure factor,  $\mathcal{S}^{\pm}(\bo)$, along the Brillouin zone paths in Fig.~\eqref{K_spec1}. The parameter are the same as in Fig.~\eqref{Tsf}.}
\label{Tsf2}
\end{figure} 
\begin{align}
\mathcal{S}^{\beta\gamma}(\bo,\omega)=-\frac{1}{\pi}\text{Im}~\mathcal{G}^{\beta\gamma}(\bo,\omega),
\label{sfa}
\end{align}
\begin{figure}[ht]
\centering
\includegraphics[width=3in]{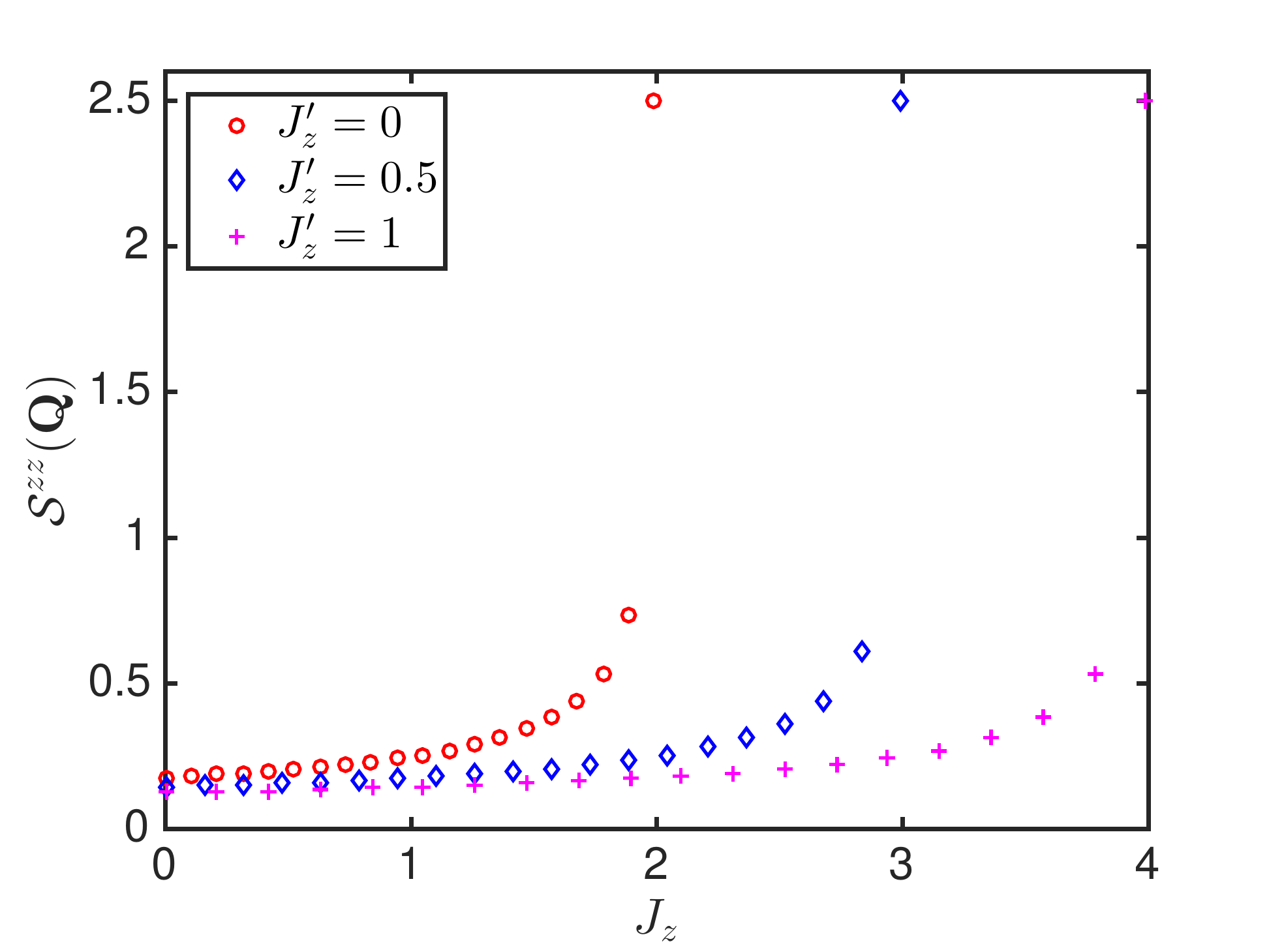}
\caption{Color online. The static structure factor,  $\mathcal{S}^{zz}(\bold{Q})$ as a function of $J_z$ at $J_{\pm\pm}=1$.}
\label{Sq}
\end{figure}
with $\mathcal{G}^{\beta\gamma}(\bo,\omega)=-i\la \mathcal{T}  S^\beta_{\bo}(t)S^\gamma_{-\bo}(0)\ra$ being the time-ordered retarded Green's function. At half-filling or zero magnetic fields, $\theta=\pi/2$, Eq.~\eqref{trans} gives $S_l^x\to S_l^{\prime z}$ and  $S_l^z\to -S_l^{\prime x}$.  The quantization axis in this case is along the $x$-axis and the off-diagonal terms are $S_{l}^\pm = S_l^z\pm iS_l^y$. Using linear spin wave theory to order $S$ we find the two static structure factors at half-filling:  $\mathcal{S}^{zz}(\bo)= S(u_\bo -v_\bo)^2/2$ and $\mathcal{S}^{\pm}(\bo)= 2S v_\bo^2$. The static structure factors  $\mathcal{S}^{zz}(\bo)$ and $\mathcal{S}^{\pm}(\bo)$  are shown in Figs.~\eqref{Tsf}  and \eqref{Tsf2} respectively. In contrast to the $U(1)$-invariant model,  the  static  structure factors show no discontinuity in the entire Brillouin zone.  Our calculation shows that  $\mathcal{S}^{zz}(\bo)$ develops sharp step-like peaks at the minima of the energy spectra and it is completely flat at the Heisenberg point $J_{\pm\pm}=J_z=1$, $J_z^\prime=0$. The off-diagonal term  $\mathcal{S}^{\pm}(\bo)$   exhibits a similar behaviour but  become completely zero except at the peaks.  At $K=\bold{Q}=(\pm 4\pi/3,0)$ we find \bea \mathcal{S}(\bold{Q})= \frac{S}{2}\sqrt{\frac{J_{\pm\pm}}{2(J_{\pm\pm}+J_z^\prime)-J_z}}.\eea
Figure \eqref{Sq} shows the plot of $\mathcal{S}(\bold{Q})$ as a function of $J_z$.

\textbf{Conclusion}--. We have explored the ground state thermodynamic properties of the easy-axis ferromagnetic phase of the quantum  triangular ice model. We observed fascinating  features which are different from the U(1)-invariant XXZ model. In particular, divergent and discontinuous quantities  in the  U(1)-invariant XXZ model are finite in the $Z_2$-invariant  XXZ model, hence all the points in the Brillouin zone contribute to the thermodynamic quantities.   Interestingly, we found that linear spin wave theory provides an accurate picture of this system with the spin-deviation operator $\la n_l\ra<0.025$ for all parameter regimes considered. In contrast to rotational invariant systems, at the Heisenberg point, the ferromagnetic state of the $Z_2$-invariant model exhibits no Goldstone mode at $\bo=0$. The excitation spectrum is gapped in the entire Brillouin zone, and the average spin-deviation is finite at low temperature near the gapped states. Hence, the $Z_2$ discrete symmetry is spontaneously broken even at finite temperature. As a result of the distinctive features of this model,  the particle density and the condensate fraction at half-filling give reasonable estimates at the level of our spin wave theory.  Also, the dynamical structure factors and the density of states exhibit interesting peaks at unusual momenta. The most distinctive feature of the $Z_2$-invariant model comes from the $\bo=0$ mode. This mode plays a very prominent role in  the unconventional phases obtained from the fully frustrated $Z_2$-invariant model.\cite{juan,sow1}  The features uncovered in this paper might be useful for experimental purposes in gapped physical systems the could be modeled with the $Z_2$-invariant model. It is also interesting to investigate the nature of this model in quantum optical lattices.\cite{mar, bec} An investigation of magnon  decay would be of interest in the $Z_2$-invariant model, this requires one to go beyond the linear spin wave approximation;  however, this is not very important at the level of our investigation since linear spin wave theory provides reasonable estimates of the thermodynamic quantities. 

\textbf{Acknowledgments}--. The author would like to thank African Institute for Mathematical Sciences (AIMS), where  this work was conducted. Research at Perimeter Institute is supported by the Government of Canada through Industry Canada and by the Province of Ontario through the Ministry of Research
and Innovation.

\end{document}